\documentstyle[editedvolume,numreferences,epsfig]{crckapb} 

\def\lsim{\lower.5ex\hbox{$\; \buildrel < \over \sim \;$}}
\def\gsim{\lower.5ex\hbox{$\; \buildrel > \over \sim \;$}}

\begin{opening}
\title{MODELING THE TIME VARIABILITY OF \protect\\
       BLACK HOLE CANDIDATES}
\subtitle{Light Curves, PSD, Lags}

\author{D. KAZANAS}
\institute{Laboratory for High Energy Astrophysics\\
           NASA/GSFC, Code 661
           Greenbelt, MD 20771}
\end{opening}

\runningtitle{MODELS OF THE TIME VARIABILITY OF BHC}

\begin{document}

% The \begin{document} command comes after the \end{opening}
% command.

\begin{abstract}
We present a model for the aperiodic variablity of accreting Black 
Hole Candidates (BHC) along with model light curves. According to the
model this variability is the  combined outcome of random (Poisson)
injection of soft photons near the center of an extended {\sl 
inhomogeneous} distribution of hot electrons (similar to those advocated
by the ADAF or ADIOS flows) and the stochastic nature of 
Compton scattering which converts these soft photons into the observed 
high energy radiation. Thus, the timing properties (PSD, lags, 
coherence) of the BHC light curves reflect, to a large extent, 
the properties of the scattering medium (which in this approximation
acts as a combination of a {\sl linear} amplifier/filter) and they can 
be used to  probe its structure, most notably the density profile of the 
scattering medium. The model accounts well for the observed PSDs and
lags and also the reduction in the RMS variability and the increase 
in the characteristic PSD frequencies with increasing source luminosity.
The electron density profiles obtained to date are consistent mainly
with those of ADIOS but also with pure ADAF flows.
\end{abstract}

\section{Introduction}

The study of the physics of accretion  powered sources, whether on 
galactic (X-ray binaries) or extragalactic systems (AGN), involves 
length scales much too small to be resolved by current technology. 
As such, this study is conducted mainly through the 
theoretical interpretation of their spectral and temporal properties. 
Until recently studies of this class of objects focused, for technical 
mainly reasons, on their X-ray spectra.  These are generally fit 
very well with those of  Comptonization of soft photons by hot 
$(T_e \lsim 10^9$ K) electrons, a process explored in great depth 
over the past twenty or so years (\cite{Suntit80}; R. Sunyaev this volume). 
Because the electron temperatures of matter 
accreting onto a black hole are expected to be similar  to those necessary 
to produce the observed spectra, it has been considered that detailed 
spectral fits of these sources would lead to insights on the dynamics of 
accretion onto the compact object. 

However, the determination of accretion dynamics requires the knowledge
of the density and size of the emitting region, neither of which is provided
by radiative transfer and spectral fitting considerations (the equations 
of radiative transfer involve the optical depth as the independent variable).
Indeed, as shown explicitly  in \cite{KHT} and \cite{HKC}, plasmas 
of very different radial extent and density profiles can yield identical 
Comptonization spectra. The degeneracy of this situation can 
lifted with the additional information provided by timing observations. 

The timing properties of BHC, however, suggest length scales 
inconsistent with the prevailing notion that the observed X-rays are 
emitted from a region of size a few Schwarzschild radii, $R_S$: 
The power spectra (PSD) of BHC exhibit most of their power at 
scales $ \sim 1 $ s, far removed from the characteristic time 
scales associated with the dynamics in the vicinity of the black 
hole horizon, $R_S/c \sim 10^{-3}$ s. Until recently rather little 
attention has been paid to this time scale discrepancy, generally 
attributed to the (unknown) mechanism ``fueling" the black hole, 
presumably operating at much larger radii; rather, more attention 
was paid to the flicker noise--type ($\propto f^{-1}$) PSDs of 
this class of sources. Novel insight into the variability properties
were introduced by \cite{miya1}, who showed that both the magnitude 
and the Fourier period dependence of the lags in the X-ray light 
curves at two different energies was inconsistent with Comptonization
by a plasma of size $\sim {\rm a ~few} R_S$; the magnitude of 
time lags ($\sim 0.1$ s) suggested an emission region of much larger
size, precluding thus an explanation of the 
observed PSDs due to a modulation of the accretion
rate onto the black hole.

These discrepancies between the expected and the observed variability 
of GBHC led \cite{KHT}, \cite{HKT}, \cite{HKC} and \cite{KH} to propose
that, contrary to the prevailing notions, the size of the 
scattering region responsible for the X-ray emission is not  $R 
\simeq 3 - 10 \; R_S$ but rather $R \gsim 10^3 R_S$, as implied by 
the PSDs and lag observations. Furthermore, the scattering medium 
(corona) is {\sl inhomogeneous}, with the electron density following 
the law, 
\begin{equation}
n(r) = \cases  {n_1 &for $r \le r_1$ \cr n_1 (r_1/r)^{p} &
for $r_2 > r > r_1$ \cr} 
\label{density}
\end{equation}
where $r$ is the radial distance from the center of the corona
(assumed  to be spherical) and $r_1$, $r_2$ are its inner and outer 
radii respectively. The index $p >0$ is a free parameter whose value 
depends on the specific dynamical model that determines the
electron density. For example, the ADAF of \cite{naryi} suggest 
$p=3/2$ and $T_e \lsim 10^9$ at radii as large as $r \simeq (m_p/m_e) 
R_S$, while models combining inflow and outflow \cite{contlov}, 
\cite{adios} (ADIOS) allow in addition  values $p \simeq 1$. As 
pointed out in \cite{KHT}, \cite{HKC} most of the data analyzed to
date suggest $p=1$ with $p = 3/2$ also acceptable in certain cases. 

\begin{figure}
\centerline{\epsfig{file=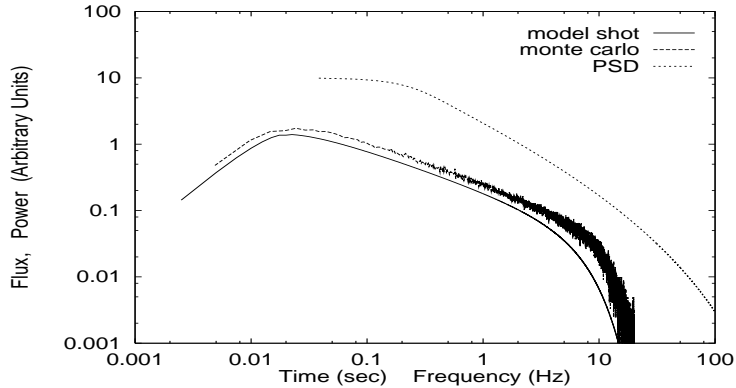,width=.80\textwidth,height=2.0in}}
%\vspace{2cm}  % amount of vertical space needed
\caption{The response function $g(t)$ of a corona with $r_1 = 10^{-2}$ 
light s, $r_2= 10$ l. s, $p = 1$, $\tau_T = 1$. The solid line is 
a fit to the MC data (dashed line). The dotted line is the Fourier 
spectrum $\vert G(\omega) \vert^2$ of $g(t)$. }
\label{shot}
\end{figure}

It is intuitively obvious that scattering in the extended configuration
of Eq. (\ref{density}) produces time lags over a range of 
Fourier periods similar to the range of radii span by the hot corona: 
Scattering at a given radius $R$ increases the X-ray energy and introduces
a lag $\Delta \,t \lsim R/c$ between the scattered and unscattered 
photons, the lag appearing at a Fourier period $P = R/c$ (\cite{HKC};
N. Kylafis these proceedings). To compare  this to the lags given in
\cite{miya1} (which are the average over all such photon pairs scattered
in a given decade in $R$, as a function of $R$) one has to multiply 
$\Delta t$ by the probability of scattering at a given radius, 
${\cal P}(R) \simeq \tau(R)$, with $\tau(R)$ the scattering depth 
over the radius $R$. For the density profile of Eq. (\ref{density}), 
$\tau(R) \propto R^{-p+1}$ and since the Fourier period  $P \propto R$, 
$\langle \Delta t \rangle \propto R^{-p+2} \propto P^{-p+2}$.  Monte 
Carlo simulations and analytical considerations have shown that 
these arguments are essentially correct \cite{HKC}. In addition, 
these simulations showed that the configuration of Eq. 
(\ref{density}) produces light curves of high coherence over the 
range $[r_2/c,r_1/c]$ of  the Fourier period, provided that the 
soft photons are injected near its center.

\section{The Model Light Curves}

Based on these considerations one can easily produce model
light curves, whose properties in the time or the Fourier 
domain can then be compared to observations for consistency.
The high coherence of the observations \cite{vaughan}
suggest injection of the soft photons at $r << r_2$. The response 
function, $g(t)$, of a corona with density profile given by Eq. 
(\ref{density}) to an instantaneous release of soft photons at 
$r\le r_1$ has been computed by a Monte Carlo simulation \cite{KHT}, 
\cite{KH}. The result of a particular case is shown in Figure \ref{shot}
along  with an analytic fit. As pointed out in \cite{HKC} these 
functions can be approximated well by a Gamma function distribution
i.e. $g(t) \propto t^{\alpha - 1}\, e^{-t/\beta}~ (0<\alpha < 
1, ~\beta >0)$.

\begin{figure}
\centerline{\epsfig{file=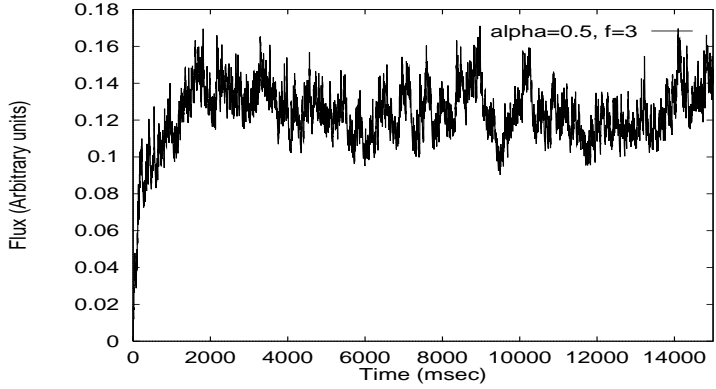,width=.80\textwidth,height=2.0in}}
%\vspace{2cm}  % amount of vertical space needed
\caption{The light curve $F(t)$ of Eq.(\ref{lightc}) for the $g(t)$ of 
Figure \ref{lc1} and $f$ = 3. }
\label{lc1}
\end{figure}

\begin{figure}
\centerline{\epsfig{file=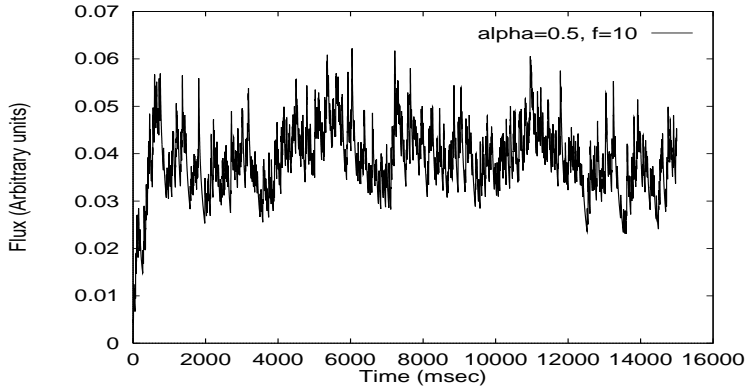,width=.80\textwidth,height=2.0in}}
%\vspace{2cm}  % amount of vertical space needed
\caption{Same as Figure \ref{lc1}, with $f = 10$. }
\label{lc2}
\end{figure}

Assuming linearity, i.e. that $T_e$ is not affected by the photon
flux, model light curves can be produced by an incoherent, random 
(Poisson) injection of shots of the form $g(t)$. The prescription 
for such a light curve is
\begin{equation}
F(t) = \sum_{i=1}^N Q_i \theta\left(t - \sum_{i=1}^N t_i
\right) g(t - t_i) ~.
\label{lightc}
\end{equation}
$Q_i$ are the shot amplitudes, assumed constand, $\theta(t - t_1)$ 
is the Heaviside function and $t_i$ is a collection of Poisson 
distributed time intervals obtained from the expression $t_i = - f 
\cdot t_0 \cdot log R_i$; $R_i$ is a random number uniformly distributed 
between 0 and 1, and $f$ is a real number, indicating the mean time 
between shots in terms of their rise time $t_0$. Figure \ref{lc1} shows
the light curve obtained using the above prescription  with  $g(t)$ 
as given in Figure \ref{shot} and $f=3$, while Figure \ref{lc2}
the light curve from an identical sequence of $R_i$'s and $f=10$.

These light curves look  similar to those of BHC sources in 
their low, hard state \cite{miya2}. Due to the Poisson nature of 
$t_i$'s and the identical shape of the shots that make up these 
light curves, their PSDs are those of an individual shot, also
shown in Figure \ref{shot} (dotted line). The form of the PSD is
also quite similar to those of BHC (see \cite{miya2}), with the
high and low frequency breaks associated respectively with the 
inner $r_1$ and outer $r_2$ edges of the scattering corona. 

\begin{figure}
\centerline{\epsfig{file=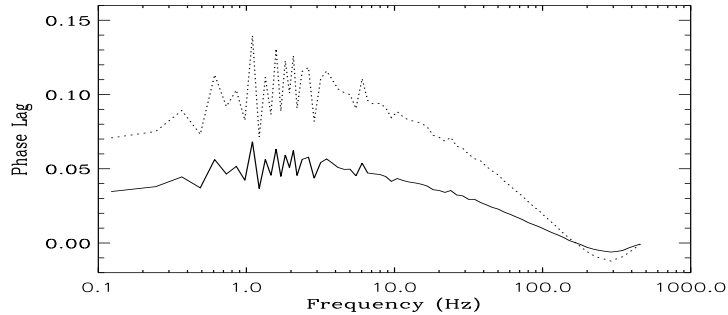,width=.90\textwidth,height=2.0in}}
%\vspace{2cm}  % amount of vertical space needed
\caption{The phase lags of two sets of model light curves with the 
following parameters (a) $\alpha =0.5$, $\beta = 16$ sec, (b) $\alpha =0.55$,
$\beta = 16$ sec, (c) $\alpha =0.6$, $\beta = 16$ sec. The two curves 
correspond to the lags between: (a) - (b) (solid line) and (a) - (c) 
(dotted line). }
\label{lags}
\end{figure}

One can easily produce, in addition, light curves for different photon 
energies, considering that for photons of a higher energy,  $g(t)$
is approximately of the same form with only a small increase in the 
parameters $\alpha$ and $\beta$ \cite{HKC}. While the resulting 
light curves are visually identical to those of  lower energies, 
their difference is 
nonetheless easily seen in the phases of the corresponding 
FFT. We have generated this way model light curves appropriate
to three different energies and then computed the corresponding 
pairs of phase lags (a)-(b), (a)-(c) as a function of the Fourier 
frequency measured in Hz; the results are shown in Figure \ref{lags}. 
The phase lags as presented in this figure are very similar to those 
of the sources analyzed in \cite{miya2} and in particular the X-ray  
transient GRO J0422+32 \cite{grove}.

A correlation between the RMS variability and the value of $f$ is 
the most prominent feature of Figures \ref{lc1} and \ref{lc2}.
Because, in the linear regime, smaller $f$ is equivalent to higher 
luminosity (more photons per unit time), this correlation provides 
a natural account of the {\sl observed} anticorrelation between the BHCs 
luminosity (in their hard spectral states) and their RMS variability.
Furthermore, consideration of this model within its natural dynamic 
framework, namely that of ADAF, provides a direct correspondence between 
the dynamical and the timing properties of these systems: In ADAF all
scales (and also $r_1$ and $r_2$) scale proportionally to $v_{ff} \,
\tau_{cool}$ with $v_{ff}$ the free-fall velocity and $\tau_{cool}$
the local cooling time. For $\tau_{cool}$ inversely proportional to 
the local denisty (as is the case with ADAF), all length scales (and 
therefore frequencies) associated with the  corona should depend on
$\dot m /\dot m_{Edd}$, decreasing with increasing  value of 
$\dot m /\dot m_{Edd}$; hence the corresponding frequencies should 
increase  with increasing luminosity, a behavior which has apparently
been observed in most accreting sources (whether neutron stars or 
black holes). 

\section{Conclusions}

(a) The model of the extended inhomogeneous corona provides
in a simple,  well understood fashion model light curves 
for BHC which have morphology, PSDs and phase (or time) lags
very  similar to those observed. (b) The dependence of the 
phase lags on Fourier period allows the determination of 
the density profile of the corona; the data are consistent
both with $p=3/2$ (ADAF) and $p=1$ (ADIOS) flows. (c) The 
dependence of the RMS variability and the PSD break frequencies
of BHC on their luminosity are also in general agreement 
with these dynamical considerations and provide a well 
defined framework within which these ideas could be tested
through more detailed comparison with observations.

{}  % Note the empty braces!

\end{document}